\documentstyle[aps,prd,floats,epsf]{revtex}

\newcommand\lsim{\mathrel{\rlap{\lower4pt\hbox{\hskip1pt$\sim$}}
     \raise1pt\hbox{$<$}}}
\newcommand\gsim{\mathrel{\rlap{\lower4pt\hbox{\hskip1pt$\sim$}}
     \raise1pt\hbox{$>$}}}
\newcommand\esim{\mathrel{\rlap{\raise2pt\hbox{\hskip0pt$\sim$}}
     \lower1pt\hbox{$-$}}}

\preprint{DAMTP-2001-26}
\date{June 13th, 2001}
\tighten
\begin{document}


\title{CMB constraints on spatial variations of
the vacuum energy density}

\author{P.\ P.\ Avelino${}^{1,2}$\thanks{
Electronic address: pedro\,@\,astro.up.pt}
A.\ Canavezes${}^{1}$\thanks{
Electronic address: alex\,@\,astro.up.pt}
J.\ P.\ M.\ de Carvalho${}^{1,3}$\thanks{
Electronic address: mauricio\,@\,astro.up.pt}
and C.\ J.\ A.\ P.\ Martins${}^{1,4,5}$\thanks{
Electronic address: C.J.A.P.Martins\,@\,damtp.cam.ac.uk}}

\address{${}^1$ Centro de Astrof\'{\i}sica, Universidade do Porto,\\
Rua das Estrelas s/n, 4150-762 Porto, Portugal}

\address{${}^2$ Dep. de F{\' \i}sica da Faculdade de Ci\^encias da
Univ. do Porto,\\ Rua do Campo Alegre 687, 4169-007 Porto, Portugal}

\address{${}^3$ Dep. de Matem\'atica Aplicada da Faculdade de Ci\^encias da
Univ. do Porto,\\ Rua das Taipas 135, 4050 Porto, Portugal}

\address{${}^4$ Department of Applied Mathematics and Theoretical Physics,\\
Centre for Mathematical Sciences, University of Cambridge\\
Wilberforce Road, Cambridge CB3 0WA, U.K.}

\address{${}^5$ Institut d'Astrophysique de Paris,\\
98 bis Boulevard Arago, 75014 Paris, France}

\maketitle
\begin{abstract}
{
In a recent article, a simple `spherical bubble' toy model for a
spatially varying vacuum energy density was introduced, and type Ia
supernovae data was used to constrain it. Here we generalize the model to
allow for the fact that we may not necessarily be at the centre of a region
with a given set of cosmological parameters, and discuss the
constraints on these models coming from Cosmic Microwave Background
Radiation data. We find tight constraints on possible spatial 
variations of the vacuum energy density for any significant deviations 
from the centre of the bubble and we comment on the relevance of our 
results.
}
\end{abstract}

\section{Introduction}
\label{secintro}

Recent Cosmic Microwave Background (CMB) \cite{cmbdata}
and type Ia supernovae \cite{snedata} data have
provided some reasonably strong evidence for an accelerating local
universe, implying that most of the `missing mass' of the universe should
be in a non-clustered form, such as a cosmological constant or quintessence.

Yet it is well known that these measurements are mostly local, so one
should be especially careful in the way they are used. In particular,
it is almost always assumed that the equation of state of the dark matter is
constant for low redshifts. However, this is  by no means well justified,
and a redshift dependence would arise if there are space and/or non-trivial 
time variations of the cosmological parameters.

Two `natural' ways in which this would happen are quintessence models
with a non-trivial equation of state, and models where a late-time
phase transition \cite{our1,our2,our3,prev} produces wall-like defects which
separate regions with different values of the cosmological parameters,
notably the matter and vacuum energy densities (and hence also the Hubble 
constant). Note that even though the motivations
for the two types of models are quite different, their observational
consequences can be fairly similar, and distinguishing between the two may 
be a non-trivial task.

In a recent paper \cite{prev}, three of the present authors have
introduced a very simple `toy model' that aims to mimic this type of
cosmological scenario with a minimal number of free parameters.
It assumes that we live inside a spherical bubble with a given set of
cosmological parameters, which is surrounded by a region where these
parameters are different. In terms of redshift, there will be a single
discrete jump on in these parameters as one goes through the wall.
A perhaps more realistic, but also definitely more
complicated toy model (in the sense of having more
free parameters) would be one with various different
`shells', and hence several jumps in the cosmological parameters at
various redshifts. Ultimately, one could get to the continuous limit where
one has the cosmological parameters varying as continuous 
functions of redshift in a non-trivial way. Something along these lines 
is discussed in \cite{multi}.

In our previous work \cite{prev}, we have used the recent measurements
of the luminosity {\em versus} redshift relation using Type Ia supernovae,
which have now been observed  out to $z\sim1.7$ \cite{newsup},
to constrain the simplest models of this type.
It was found that presently available observations are only
constraining at very low redshifts $z \lsim 0.5$ (as was to be expected),
but it was also independently confirmed that the high red-shift supernovae
data does prefer a relatively large positive cosmological constant.

Here we will discuss the constraints on this type of models coming from
CMB data, while also allowing for a further effect. Indeed, we
allow for the possibility that we are not at the centre of the inner domain
or, in other words, that there is a further anisotropic component
in the CMB. Obviously the observed near-isotropy of the CMB will impose
quite strong constraints on our position, but it is still important to
determine how much freedom is allowed if one considers that part of
the observed anisotropy comes from such a displacement.

In the next section we briefly review our toy model and discuss the
evolution of cosmological perturbations within its framework.
In Sect. III we show, mainly as an amusing interlude,
how one could {\em design} a `mimic model' of this kind which would
reproduce the CMB spectrum predicted by standard paradigms such as
inflation. Our main results are presented
and discussed in Sect. IV, and we conclude in Sect. V.

\section{Cosmological perturbations in a spherical bubble model}
\label{secmodel}

We shall again consider a simplified model in which the universe consists
of a spherically symmetric region (domain) with a given set of
cosmological parameters, which is surrounded by another region where
the values of the cosmological parameters (in particular, that of
the vacuum energy density) are different. We shall refer to these
two values of the vacuum energy density as
$V_-$ and $V_+$ respectively. Note that the cosmological parameters
will in general be different in both regions, but their variations
are not totally independent---they are such as to make the universe
flat both inside and outside the domain (this point is discussed in
detail in \cite{prev}).

As in our previous work \cite{prev}, we assume that the red-shift of the 
domain wall, as measured by an observer at the centre of the inner
spherical domain, is $z_*$. However, we now allow for the possibility
that we are not at the centre of the domain. Specifically, we assume
that we live a red-shift $z_\Delta \le z_*$ away from this
`central' observer.

Again, we are assuming that the thin region separating the two domains
considered (domain wall) does not generate relevant CMB fluctuations.
This happens if the potential of the field is small enough at the
origin \cite{prev}.
We shall also assume that the domain walls are frozen in comoving 
coordinates, which again will be a good approximation provided that
friction is important \cite{mod}.

We shall parametrize the vacuum energy density by
\begin{equation}
\Omega_\Lambda \equiv \frac{V_-}{\rho_c},
\label{defol}
\end{equation}
where $\rho_c$ is the critical density,
and we define $\Delta \Omega_\Lambda$ as
\begin{equation}
\Delta \Omega_\Lambda (r) = \frac{\delta \rho_\Lambda(r)}{\rho_c} =
\frac{\rho_\Lambda(r) - V_-}{\rho_c},
\label{defdol}
\end{equation}
where $\rho_\Lambda(r)$ is the vacuum energy density
at the point in question. Hence, this can have two possible
values:  $0$ if we are inside the inner region,
and $(V_+-V_-)/\rho_c$ in the outer domain.

In the conformal-Newtonian gauge, the line-element for a flat
Friedmann-Robertson-Walker background and scalar metric perturbations
can be written as
\begin{equation}
ds^2 = a^2(\eta) \left[(1+2 \Phi) c^2 d \eta^2 - (1-2 \Phi) (dr^2 +
r^2 d
\theta^2 + r^2 \sin^2 \theta d \phi^2) \right],
\label{metric}
\end{equation}
assuming that the anisotropic stresses are small. Here, $\Phi$ is the metric
perturbation, $c$ is the speed of light in vacuum, $a$ is the
scale factor, $\eta$ is the conformal time, and $r$,
$\theta$ and $\phi$ are spatial coordinates.

Given that the vacuum energy becomes dominant
only for recent epochs we shall be concerned with
the evolution of perturbations only in the matter-dominated era,
thus neglecting the contribution of the radiation component.
The evolution of the scale factor $a$ is governed by the
Friedmann equation\footnote{A dot denotes a derivative with respect to
conformal time, ${\cal H} ={\dot a} / a$, the index `0'
means that the quantities are to be evaluated at the present time and
we have taken $a_0 = 1$ and ${\cal H}_0 = 1$ (so that the conformal time is
measured in units of ${\cal H}^{-1}_0$).}, which can be written as
\begin{equation} {\cal H}^2  = {{\Omega}}_m^0 a^{-1} +
{\Omega}_\Lambda^0 a^2\, .
\label{friedmann}
\end{equation}
Note that the background matter and vacuum energy densities at an
arbitrary epoch can be written as
\begin{equation} {\Omega}_m=\frac{{{\Omega}}_m^0}{{{\Omega}}_m^0 +
{\Omega}_\Lambda^0 a^3}
\label{omt}
\end{equation}
and ${\Omega}_\Lambda = 1-{\Omega}_m$.

The linear evolution equation of the scalar perturbations is given by
(see for example \cite{mfb})
\begin{equation}
    \ddot \Phi + 3 {\cal H} \dot \Phi  +
[2 {\dot {\cal H}} + {\cal H}^2]\Phi = 4 \pi G a^2 \delta p =
- {3 \over 2} a^2 \Delta \Omega_\Lambda^0 ,
     \label{scalar}
\end{equation}
where $\delta p$ is the pressure perturbation and
\begin{equation}
2 {\dot {\cal H}}  = - { {\Omega}}_m^0 a^{-1} +
2 {\Omega}_\Lambda^0 a^2.
\label{doth}
\end{equation}
Given that the source term in the outer domain
($\propto a^2 \Delta \Omega_\Lambda^0$) is
only important near the present time, we assume the
following initial conditions for eq.~(\ref{scalar}):
\begin{equation}
\Phi(0)=0, \qquad {\dot \Phi}(0)=0,
\end{equation}
both in the inner and the outer regions. We note that since the source term
is absent in the inner region, the metric perturbation is always zero there.

The novel feature of this type of model is that
in the presence of the scalar metric
perturbations defined by eqn. (\ref{scalar}) there is a shift
in the CMB temperature given by
\begin{equation}
\frac{\Delta T}{T} = -2\Phi_+.
\label{dtot}
\end{equation}
where $\Phi_+$ is the value of $\Phi$ in the
outer region at the time when the light crossed the domain wall (recall
that here we are assuming that $\Phi=0$ in the inner region).
This temperature shift is due to the relative velocity between comoving
observers on either side of the domain wall. 

In order to compute the spectrum of CMB fluctuations generated in this model 
as a function of its parameters (given the values of 
$\Omega_m^0$ and $\Omega_\Lambda^0$) 
we need to calculate the time when the CMB radiation that reaches us today
crossed the domain wall. This, of course, depends on the direction we are 
looking at if we're not at the centre of the inner bubble.

We will parametrize this direction by the 
angle $\theta={\bf n} \cdot {\bf n}_{cl}$ where the unit vector 
${\bf n}$ defines an arbitrary direction on the sky and 
${\bf n}_{cl}\equiv{\bf r}/|{\bf r}|$ where ${\bf r}$ is the vector 
that points from the observer on earth 
to the closest point on the domain wall. Using the Friedmann equation it is 
straightforward to compute the comoving distances $d(z_*)$ and $d(z_\Delta)$ 
corresponding to the red-shifts $z_*$ and $z_\Delta$,
\begin{equation}
d(z)= c \int_0^z {{dz'} \over { \left[ \Omega^0_m (1+z')^3 +
\Omega^0_\Lambda\right]^{1/2}}}. 
\label{cdis}
\end{equation}
We find it useful to parametrize our position within the bubble by
\begin{equation}
\epsilon \equiv \frac{d(z_\Delta)}{d(z_*)} \,,
\label{epsilon}
\end{equation}
that is, the ratio of our comoving distance to the bubble centre and
the comoving bubble radius.
On the other hand, the comoving distance to the domain wall as seen
by us as a function of the angle $\theta$ is given by
\begin{equation}
D(\theta) = {\sqrt{(d(z_*)\cos\theta
- d(z_\Delta))^2 +(d(z_*) \sin \theta)^2}}, 
\label{dtheta}
\end{equation}
and the conformal time when the CMB radiation that reaches us today
crossed the domain wall at that particular direction is then simply
given by
\begin{equation}
\eta(\theta) = \eta_0 - \frac{D(\theta)}{c}\,.
\label{etat}
\end{equation}
The value of $\Phi_+ (\theta) \equiv \Phi(\eta)$ can now be calculated using 
eqn. (\ref{scalar}).

Before moving on to discuss the results of this analysis in
Sect. IV, we make a brief detour to discuss an interesting property of
this type of models.

\section{A mimic model}
\label{mimic}

Let us assume for the sake of simplicity that $\Omega_m=1$ and 
$\Omega_\Lambda=0$ such that eq.~(\ref{scalar}) becomes
\begin{equation}
\ddot \Phi +  \frac{6}{\eta} \dot \Phi  = 
\eta^{-6} \frac{\partial (\eta^6 \dot \Phi)}{\partial \eta} 
= -\frac{3}{8} \eta^2 
\Delta \Omega_\Lambda^0\,,
\label{scalm}
\end{equation}
where we have made use of the fact that with our conventions one
has $a=\eta^2/\eta_0^2$ and $\eta_0=2 
{\cal H}^{-1}_0 = 2$. Solving this equation with the boundary conditions 
\begin{equation}
\Phi(0)=0, \qquad {\dot \Phi}(0)=0,
\end{equation}
in the outer region we obtain
\begin{equation}
\Phi(\eta) = -\frac{1}{96} \eta^4 \Delta \Omega_\Lambda^0 = -\frac{1}{6}
(1+z)^{-2}  \Delta \Omega_\Lambda^0.
\label{scals}
\end{equation}
Let us consider a domain wall which is at a red-shift $z$ from us. Consider 
small spatial fluctuations in the red-shift of the domain wall parametrized 
by $\Delta z(\theta,\phi)$ where $\theta$ and $\phi$ are angular coordinates. 
These fluctuations will induce CMB perturbations with
\begin{equation}
\frac{\Delta T}{T} = -2 \Delta \Phi_+ =  -\frac{2}{3}
\Delta z (1+z)^{-3}  \Delta \Omega_\Lambda^0.
\label{dtot1}
\end{equation}
where we have subtracted the multipole of order $0$ of the CMB fluctuations.
This has been done for a flat matter dominated background but it is obvious
to see that it can be done more generally.

This demonstrates that it would be 
possible in general to generate any spectrum of CMB fluctuations by an 
appropriate design of the domain wall. It would, of course, be extremely 
unlikely that a domain wall would have the precise shape necessary to 
produce the CMB spectrum predicted in popular models for the origin of CMB and 
large scale structure (e.g. inflation). However, it would not be so 
implausible that this model could have some kind of impact on 
the CMB fluctuations. 

Irrespective of the likelihood of this model, we think it serves to highlight
an important point, namely that a good test to the predictions 
of current models for structure formation consists in studying to what 
extent these could be produced by other models (even if they might seem 
unlikely in the light of our current theoretical prejudices).

\section{Results and discussion}
\label{results}

\begin{figure}
\vbox{\centerline{
\epsfxsize=0.5\hsize\epsfbox{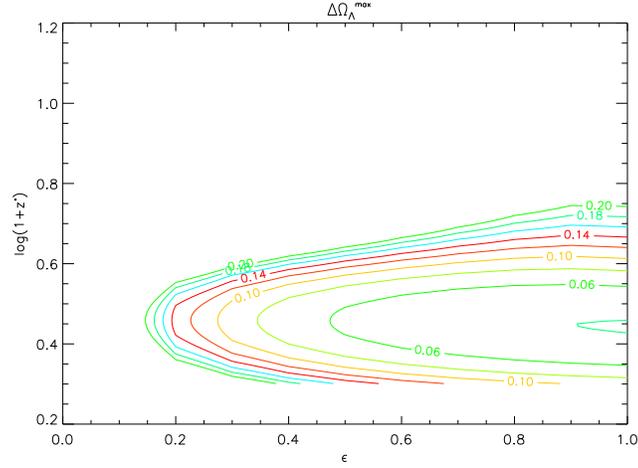}}
\vskip.4in}
\caption{The maximum variation in the vacuum energy density with respect to
the present local value (parametrized by $\Delta \Omega_\Lambda^{\rm max}$) 
allowed by the COBE dipole, as a function of our 
position within the spherical bubble. This position is parametrized by the
redshift, $z_*$, of the spherical wall as measured by an observer at its centre,
and by the ratio, $\epsilon$, of our comoving distance to the bubble centre and
the comoving bubble radius.}
\label{fig1}
\end{figure}

\begin{figure}
\vbox{\centerline{
\epsfxsize=0.5\hsize\epsfbox{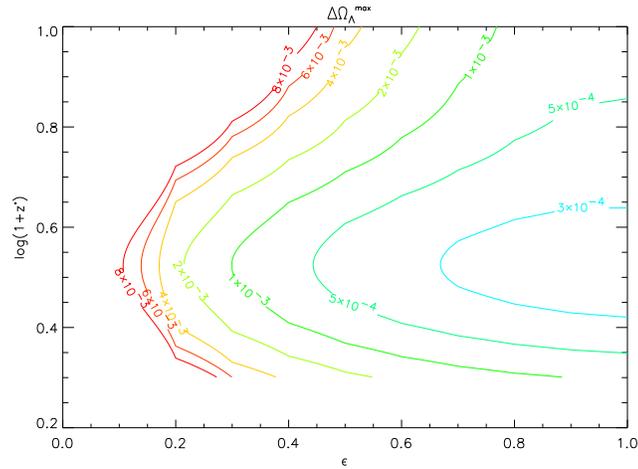}}
\vskip.4in}
\caption{The maximum variation in the vacuum energy density with respect to
the present local value (parametrized by $\Delta \Omega_\Lambda^{\rm max}$) 
allowed by the COBE quadrupole , as a function of our
position within the spherical bubble. This position is parametrized by the
redshift, $z_*$, of the spherical wall as measured by an observer at its centre,
and by the ratio, $\epsilon$, of our comoving distance to the bubble centre and
the comoving bubble radius.}
\label{fig2}
\end{figure}

Our results can be conveniently summarized by figs. \ref{fig1} and \ref{fig2}.
We have assumed that the present local universe is characterized
by cosmological densities $\Omega_m^0=0.3$ and $\Omega_\Lambda^0=0.7$,
and calculated the maximum value allowed by currently existing
observations for $\Delta \Omega_\Lambda^0$ in the outer domain 
($\Delta \Omega_\Lambda^{\rm max}$) as a function of the parameters 
$\epsilon$ and $z_*$. 
Figs. \ref{fig1} and \ref{fig2}
show the result of this calculation for the dipole and quadrupole measured by 
COBE \cite{bennett}.

As expected, there are no direct CMB constraints, to first order,
if we happen to be at the centre of the domain, and they become
progressively tighter as we move away from this position. We note that
if the observer is right at the centre of the spherical domain,
a further (second-order) contribution would come from the integrated
Sachs-Wolfe effect. Being second order, this would be subdominant
everywhere except extremely close to the centre ($\epsilon \sim
0$), and in this limit the supernovae constraints discussed in
\cite{prev} are currently much stronger. For this reason we have
ignored this effect in the present analysis, although it should be
considered in the future when more precise data becomes available.

Also as expected, there are no constraints
in the limits of an infinitely small ($z_*\rightarrow 0$) or infinitely
large ($z_*\rightarrow \infty$) inner region. This means that for any given
$\epsilon$ there will be some intermediate value of the redshift for which the
the allowed variation in the vacuum energy density is minimal. In fact it
turns out that, at least to a first approximation,
this particular redshift is independent of our position
within the bubble (which is measured by $\epsilon$). For the currently
favoured cosmological model, with $\Omega_m^0=0.3$ and
$\Omega_\Lambda^0=0.7$, this redshift is found to be $z\sim2.2$. We also point 
out that, the quadrupole constraints (Fig 2)  are, in general, 
much stronger than the dipole constraints (Fig 1). This is due to the much 
larger amplitude of the observed dipole CMB anisotropy compared with the 
quadrupole.

Several models attempting to reconstruct the matter distribution from 
the observed velocity field have been proposed. However, even the best 
ones are left with an unexplained relatively high residual velocity 
of $\sim 200 km\,s^{-1}$, too large to be an unknown motion of our 
galaxy with respect to the Local Group reference frame 
(see for example \cite{Tonry,mrrdip}).
The possibility that this discrepancy 
might be explained by a model such as the `bubble' model presented in 
this paper is, however, unfavoured by our results. The relatively large  
values of $\Delta \Omega_\Lambda^{\rm max}$ allowed by the COBE dipole 
make this possibility rather unlikely.

In any case, our analysis confirms the intuitive expectation that the CMB
constraints on anisotropies of this kind are extremely tight, or in
other words that if such a model is a realistic approximation to the
evolution of the universe, then we must be very close to the centre
of the domain. Still another way of saying it is that, if non-trivial
variations of the cosmological parameters as a function of the redshift
did occur in the past, they are more likely to result from scenarios
where the redshift variations are the result of time variations which
maintain spatial homogeneity or isotropy (such as quintessence models, for
example), since in scenarios where there are also spatial variations
we are forced by current observations to occupy a fairly privileged
(and arguably unlikely) position.

\section{Conclusions}
\label{conclusions}

In an earlier work we have introduced a very simple `spherical bubble' toy
model for a spatially varying vacuum energy density, and type Ia
supernovae data was used to constrain it. Here we have generalized this
model, in order to account for the possibility
that we may not be at the centre of a region
with a given set of cosmological parameters. We have then discussed the
constraints on this model coming from CMBR data. As expected, any
significant deviations from isotropy are tightly constrained.

Ultimately, our toy model aims to be a simple mimic for non-trivial
variations of cosmological parameters. As we have pointed out before,
these variations as one probes higher redshifts could result either from
genuine time-variations (an example of which would be quintessence
models with a non-trivial equation of state) or from spatial
variations (an example of which would be models where late-time
phase transitions produce a domain-like structure). 

The present work, together with \cite{prev} quantitatively confirms the
intuitive expectation that models with spatial variations are much more
constrained, in the sense that the near-isotropy of the CMBR  imposes
quite strong restrictions on our position within such a domain, whereas
no such constraints exist in the case of genuine time variations.
Obviously, our toy model is rather simplified, and a more detailed
analysis is required to constrain specific models. We shall return to
this interesting topic in a future publication.

\acknowledgements

C. M. is grateful for the hospitality at IAP and DARC (Observatoire de
Paris-Meudon), where this work was completed.
A. C. and C. M. are funded by FCT (Portugal) under ``Programa PRAXIS XXI''
(grant nos.\ PRAXISXXI/BPD/18826/98 and FMRH/BPD/1600/2000 respectively).
We thank ``Centro de Astrof{\' \i}sica da Universidade do Porto''
(CAUP) for the facilities  provided.



\begin{references}
\bibitem{cmbdata}
C.B. Netterfield {\em et al.}, astro-ph/0104460 (2001);

C. Pryke {\em et al.}, astro-ph/0104490 (2001);

R. Stompor {\em et al.}, astro-ph/0105062 (2001).
\bibitem{snedata}
A. G. Riess {\em et al.}, Astron. J. {\bf 116}, 1009 (1998);

P. M. Garnavich {\em et al.}, Ap. J. Lett. {\bf 493}, L53 (1998);

S. Perlmutter {\em et al.}, Ap. J. {\bf 517}, 465 (1999).
\bibitem{our1}
P. P. Avelino, J. P. M de Carvalho and C. J. A. P. Martins, Phys. Lett.
{\bf B501}, 257 (2001).
\bibitem{our2}
P.P. Avelino and C.J.A.P. Martins, Phys. Rev. {\bf D62} (2000),
103510 (2000).
\bibitem{our3}
P.P. Avelino, J.P.M. de Carvalho, C.J.A.P. Martins and J.C.R.E.
Oliveira, astro-ph/0004227 (to appear in Phys. Lett. B).
\bibitem{prev}
P.P. Avelino, J P.M. de Carvalho and C.J.A.P. Martins, astro-ph/0103075
(to appear in Phys. Rev. D).
\bibitem{multi}
Y. Wang and P.M. Garnavich, astro-ph/0101040 (2001);

M. Tegmark, astro-ph/0101354 (2001).
\bibitem{newsup}
A.G. Riess {\em et al.}, astro-ph/0104455 (2001).
\bibitem{mod}
C.J.A.P. Martins and E.P.S. Shellard, hep-ph/0003298 (2000).
\bibitem{mfb}
V.F. Mukhanov, H.A. Feldman and R.H. Brandenberger,
Phys. Rep. {\bf 215}, 203 (1992).
\bibitem{bennett}
C.L. Bennett {\em et al}, Ap. J. Lett. {\bf 464},1 (1996).
\bibitem{Tonry}
J.L.\ Tonry {\em et al.}, astro-ph/9907062 (1999).
\bibitem{mrrdip}
M. Rowan-Robinson {\em et al.}, MNRAS, in press (see also astro-ph/9912223).
\end{references}
\end{document}